\documentclass[a4paper,12pt,reqno,superscriptaddress,nofootinbib]{revtex4}
\usepackage[centertags]{amsmath}
\usepackage{amsfonts}
\usepackage{amssymb}
\usepackage{amsthm}
\usepackage{newlfont}
\usepackage{stmaryrd}
\usepackage{mathrsfs}
\usepackage{euscript}
\usepackage{siunitx}
\usepackage{physics}
\usepackage{algorithm2e}
\usepackage{graphicx,subcaption}
\usepackage{enumitem}
\usepackage{natbib}
\usepackage{graphicx}
\usepackage{color}
\usepackage{floatrow}
\usepackage{caption}
\usepackage{hyperref}
\usepackage{hhline}
\usepackage{hyperref}
\usepackage{cleveref}
\usepackage{import}

\theoremstyle{plain}

\theoremstyle{definition}

\theoremstyle{remark}

\numberwithin{equation}{section}

\let\ve=\varepsilon

\newcommand{\opunit}{\text{1}\kern-0.22em\text{l}}

\DeclareMathAlphabet{\mathpzc}{OT1}{pzc}{m}{it}

\newcommand{\id}{\textrm{d}}

\usepackage{floatrow}
\usepackage{caption}

\usepackage{color,soul}

\newcommand{\tm}{\textcolor{magenta}}

\DeclareCaptionJustification{justified}{\leftskip=0pt \rightskip=0pt \parfillskip=0pt plus 1fil}
\begin{document}
\title{Heat capacity \\of periodically driven two--level systems}
\author{Elena Rufeil Fiori$^{1,2}$ and Christian Maes$^1$\\ {\it$^1$Department of Physics and Astronomy, KU Leuven, Belgium.}\\{\it$^2$Facultad de Matemática, Astronomía, Física y Computación, CONICET--Universidad Nacional de C\'ordoba, Argentina.}}

\keywords{nonequilibrium calorimetry, two-level systems}
\begin{abstract}
    We define the heat capacity for steady periodically driven systems and as an example we compute it for dissipative two--level systems where the energy gap is time--modulated. There, as a function of ambient temperature, the Schottky peak remains the dominant feature. Yet, in contrast with equilibrium, the quasistatic thermal response of a nonequilibrium system also reveals kinetic information present in the transition rates; {\it e.g.}, the heat capacity depends on the time--symmetric reactivities and changes by the presence of a kinetic barrier. It still vanishes though at absolute zero, in accord with an extended Nernst heat postulate, but at a different rate from the equilibrium case. More generally, we discuss the dependence on driving frequency and amplitude.  
\end{abstract}

\maketitle

\section{Introduction}
Over the last decades, dissipative aspects of driven and active systems are intensively researched in condensed matter and chemical physics laboratories, \cite{pop, wang, epl, jir, subas, dm, ast, ms2, dio}. Nonequilibrium calorimetry is of increasing interest there, to investigate how heat reveals not only energetic but also kinetic and functional characteristics of natural and artificial materials. 
There are multiple ways for a system to be out of equilibrium. The system may be transient, relaxing possibly slowly but ultimately to equilibrium as is typical for glasses and systems with important kinetic constraints. The system may be bulk--driven, {\it e.g.}, by the application of rotational forces, or boundary--driven, from having different pressures or chemical potentials at its boundary. The system may also be active as being in contact with internal degrees of freedom that fuel motion. Finally, the system may also be subject to time--periodic forces so that asymptotically there appears a steady time--dependent condition which is not stationary but periodic. All the same, that steady system will dissipate heat in a thermal environment when coupled to it. In fact, the thermal properties of such {\it steady time--dependent} systems are less explored theoretically than their stationary (time--independent but bulk or boundary driven) counterparts. Nevertheless, driving at nonzero frequency is a very common realization of a nonequilibrium condition, especially in experimental and computational setups, \cite{ND,cera,dio}. For a variety of science--of--life and engineering purposes, it is essential to understand how heat capacities and conductivities are affected by temporal driving. On top, it obviously constitutes a foundational question in constructing nonequilibrium statistical mechanics as well, \cite{jchemphys}.\\

In the present paper, we define the corresponding nonequilibrium heat capacity and we illustrate the procedure and general characteristics with the simplest of systems, a dissipative two--level system where the energy difference is varying periodically in time. Two--level systems (2LS) are ubiquitous ingredients of a great variety of materials, as represented by magnetic spins, dipoles, or molecules in contact with crystal phonons. Yet, as such, a 2LS cannot be bulk-- or boundary--driven. Still, they are easily driven by, {\it e.g.}, time--dependent variations of the energy gap and  kinetic barriers. In fact, artificial double--well systems imitate 2LS and are playing an increasing role as nonequilibrium models in room-temperature experimental soft matter physics, \cite{ignacio2013, ignacio2014}, even apart from their relevance for solid--state applications at low temperatures.
Moreover, as will be easily understood, the methodology outlined below for steady periodically--driven 2LS can easily be generalized to multilevel systems.\\ 

For the general theoretical background about thermal response and heat capacities of stationary nonequilibrium systems, we refer to \cite{calo, epl, scr}. In particular, for nonequilibrium, heat capacity is defined in terms of excess heat, and there is no direct relation with entropy or entropy production, except via the notion of expected dissipated power. As usual in thermodynamics, a quasistatic setup is needed and no higher--order corrections are considered than those needed for characterize the excess dissipated power (to be explained in the next section). On the other hand, the most promising experimental setup is presented in Fig.~\ref{fig_scheme} and follows the scheme of AC--calorimetry, explained in the next section. A version of it was used before for nonequilibrium calorimetric experiment in \cite{cera}.\\

The paper starts in the next section with an operational description of thermal response for periodically driven systems.  It corresponds to AC--measurements of the heat capacity for temporally driven systems, that builds on recent work for defining and measuring nonequilibrium heat capacities for stationary systems \cite{cal}. The main idea remains to identify the excess heat released by quasistatic variations in the temperature of a heat bath with which the system is in thermal contact. In the present scenario, the steady dissipated power before and (long) after a small temperature change are both oscillating in time, making the notion of excess heat more problematic.\\   
In Sections \ref{2level} and \ref{barrier} we introduce the concrete time--dependent dynamics we are dealing with. They are described by time--periodic two--state Markov processes. We present the main observations on their heat capacities as a function of temperature, driving, and kinetic parameters. They all show the typical Schottky anomaly but there are significant changes when compared with the equilibrium case. In particular, while the heat capacity still vanishes at very low temperatures, its decay rate is slower in the nonequilibrium steady state with respect to the equilibrium case. Moreover, the heat capacity detects the presence of a kinetic barrier between the two levels, which is invisible in the equilibrium heat capacity. We also give the dependence on driving frequency and amplitude. 

\section{AC--calorimetry}

Heat capacity, whether for equilibrium \cite{bookcal, nanocal} or nonequilibrium \cite{dio} systems, measures the change in heat as a direct response to a slowly changing environmental temperature, while a choice of constraints is being specified. We refer to \cite{jchemphys} for a motivation and justification concerning nonequilibrium calorimetry. 
As a reference, we suppose an open equilibrium system in contact with a heat bath at inverse temperature $\beta$. It becomes a nonequilibrium system when work is being done on the system, periodically varying some external parameters with frequency $\omega_0$, such as external fields, pressure, or volume as makes the case. We assume that after some initial transient, the system reaches a steady condition with periodic variations in its observables. The question we address in the present paper is to identify the response to a small temperature variation in the heat bath, {\it i.e.}, to define the heat capacity as a function of $\beta, \omega_0$ and of further system parameters such as the amplitude of the work or the values of energy differences in the reference system.

\begin{figure}[ht!]
    \centering
    \includegraphics[width=0.4\textwidth]{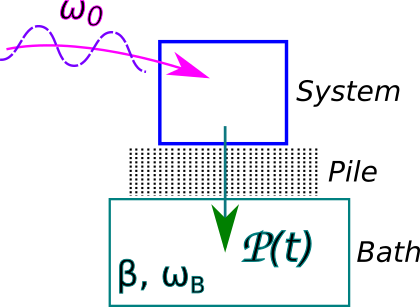}
    \caption{\underline{Sketch of the most promising calorimetric setup}. The open system or sample is acted upon by an external periodic variation of system parameters. It reaches a steady condition in contact with a heat bath at inverse temperature $\beta$ to which the work is dissipated as heat. When the temperature of the heat bath is slowly varied, say with small frequency $\omega_\text{B}$, we get a heat flux ${\cal P}(t)$ which can be measured via the thermopile connecting the sample with the bath.}
\label{fig_scheme}
\end{figure}

We refer to Fig.~\ref{fig_scheme} for a sketch of the procedure which is experimentally feasible and has indeed been used for pioneering calorimetry in nonequilibrium systems, \cite{cera, cerro}. For other experimental work on AC--calorimetry, we refer to \cite{aclowt}. There is also a clear theoretical model behind it, allowing detailed calculations, which we will now explain.

Remember that work is delivered to the system, which is constantly dissipated in the heat bath. We can measure the heat flux or dissipated power ${\cal P}^0(t)$ to that bath {\it e.g.} via a thermopile that converts heat into electric work. ${\cal P}^0(t)$ is periodic with frequency $\omega_0$. Next, to get to the thermal response, we slowly vary the bath temperature $\beta^{-1}$, with
\begin{equation}\label{timebeta}
\beta_t = \beta \,(1+ \varepsilon \sin(\omega_\text{B} t)),
\end{equation}
using a small frequency $\omega_\text{B} \ll \omega_0$ and small amplitude $\ve$, allowing the quasistatic regime for thermal response.  There is now a new (total) dissipated power ${\cal P}(t)$, see  Fig.~\ref{fig_scheme}, in which the time--dependence of the temperature enters as well:
\begin{equation}
{\cal P}(t) = {\cal P}^0(t) + \ve \,{\cal P}^1(t) + O(\ve^2), \nonumber
\end{equation}
and ${\cal P}^1(t)$ depends on $\omega_\text{B}$. Note that ${\cal P}(t)$ need not be periodic, except when $\omega_0$ and $\omega_\text{B}$ are commensurable.\\
The thermal response is in ${\cal P}^1(t)$.  The heat capacity $C=C(\beta, \omega_0,\ldots)$ is obtained from understanding how the power $\cal P$ depends on the rate of temperature--variation: with $k_B=1$, as $\omega_\text{B}\rightarrow 0$, and $\ve\rightarrow 0$,
\begin{equation}
  C =\beta_t^2\,\frac{\partial {\cal P}}{\partial(\id \beta_t/\id t)} =  \frac{\beta}{\ve\omega_\text{B}}\, \frac{\partial {\cal P}}{\partial( \cos\omega_\text{B} t)}.
\end{equation}
More specifically, by following the logic of AC--calorimetry as exposed in Ref.\cite{cal} for stationary nonequilibrium systems, we get the heat capacity $C$ as the out--of--phase component in minus the excess heat current ${\cal P}^1$: 
\begin{equation}\label{eq_C}
  C 
  =\frac{\beta}{\pi}\, \int_{0}^{2\pi/\omega_\text{B}} {\cal P}^{1}(t)\, \cos(\omega_\text{B} t)\, \id t, \qquad \text{ as  }\; \omega_\text{B}\rightarrow 0, \,\ve\rightarrow 0
\end{equation}

In the following sections, we illustrate the procedure for a two--level system and discuss the results, and how they illuminate the kinetics.  

\section{Time--dependent two--level systems}\label{2level}

\subsection{Setup}
\begin{figure}[ht!]
    \centering
    \includegraphics[width=0.4\textwidth]{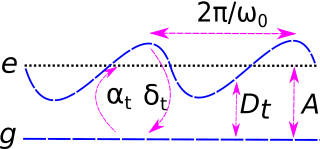}
    \caption{Two--level system with a time--periodic energy difference $D_t$ between the excited state $e$ and the ground state $g$. The variation of $D_t$ is periodic in time (with frequency $\omega_0$) but does not need to be harmonic around the reference value $A$. Transition rates are denoted by $\alpha_t$ and $\delta_t$ to move up and down in energy, respectively.}
\label{fig_harm_scheme}
\end{figure}

We consider a two--level system where we fix the ground state energy (to be zero) while we vary the energy $D_t>0$ of the excited level; see Fig.~\ref{fig_harm_scheme}.  Possible scenarios include time-dependent volumes or magnetic fields. 

 At fixed inverse temperature $\beta$ of the surrounding bath, transition rates between the ground state $g$ and the excited state $e$ are parameterized as
\begin{equation}\label{rates}
    k_t^0(g,e) = \nu_t^0\,e^{-\beta D_t/2},\qquad  k_t^0(e,g) = \nu_t^0\,e^{\beta D_t/2},
\end{equation}
with  $\nu_t^0$ a possibly time--dependent reactivity. We assume that the rates $k_t$ are periodic in time with frequency $\omega_0$.\\
The heat $q_t$ given to the bath during the jump $g\rightarrow e$ equals $q_t(g,e) = -D_t$, and for $e\rightarrow g$ it is $q_t(e,g) = D_t$. 
The expected power dissipated to the heat bath (at constant temperature) is therefore
\begin{equation}\label{power}
{\cal P}^0_t = -D_t  k_t^0(g,e) \rho_t^0(g) +  D_t  k_t^0(e,g) \rho_t^0(e),
\end{equation}
where $\rho_t^0$ is the level probability at time $t$ (depending on the initial condition), satisfying the time--dependent Master equation,
\begin{equation}
\partial_t \rho_t^0(e) =  k_t^0(g,e) \rho_t^0(g) - k_t^0(e,g) \rho_t^0(e).
\end{equation}
That is the standard setup for dissipative two--level systems in the incoherent approximation. Asymptotically in time, the system reaches a steady nonequilibrium condition where the occupation probabilities $\rho_t^0$ vary periodically with frequency $\omega_0$. That process defines the nonequilibrium system for which we want to determine the heat capacity.\\

The definition (and suggested experimental measurement method) \eqref{eq_C} of the heat capacity quantifies how the power \eqref{power} gets modified for small temperature variations around $\beta^{-1}$.  We apply the same definition of expected heat flux as in \eqref{power} but now we use \eqref{timebeta} to replace $\beta\rightarrow \beta_t$ in \eqref{rates},  changing $k_t^0\rightarrow k_t$: 
\begin{equation}\label{power1}
{\cal P}_t = -D_t\,  k_t(g,e) \rho_t(g) +  D_t\,  k_t(e,g) \rho_t(e),
\end{equation}
with 
\begin{equation}\label{me}
\partial_t \rho_t(e) =  k_t(g,e) \rho_t(g) - k_t(e,g) \rho_t(e).
\end{equation}
Applying the time-dependent inverse temperature $\beta_t$  in \eqref{timebeta}, for the occupation probabilities we  find in general
\begin{equation}\label{rh}
\rho_t =\rho_t^0 + \varepsilon \rho_t^1 + O(\ve^2,e^{-\gamma t})
\end{equation}
for relaxation rate $\gamma>0$ which we assume is strictly positive, and with small variation amplitude $\ve$.  Then, similarly, from \eqref{power1} we get
\begin{equation}\label{power2}
{\cal P}_t = {\cal P}_t^0 + \ve\, {\cal P}_t^1 + O(\ve^2,e^{-\gamma t}),
\end{equation}
and the excess power ${\cal P}_t^1$ is used in \eqref{eq_C} to get the heat capacity.\\

We make the above explicit for a number of cases and start with a harmonically varying energy gap $D_t$, followed by a stroboscopic time--variation.  We deal with the presence of a kinetic barrier in Section \ref{barrier}. The logic is the same in each case.

\subsection{Harmonically varying energy}
\label{harmo} 

\begin{figure}[h]
    \centering
    \includegraphics[width=0.6\textwidth]{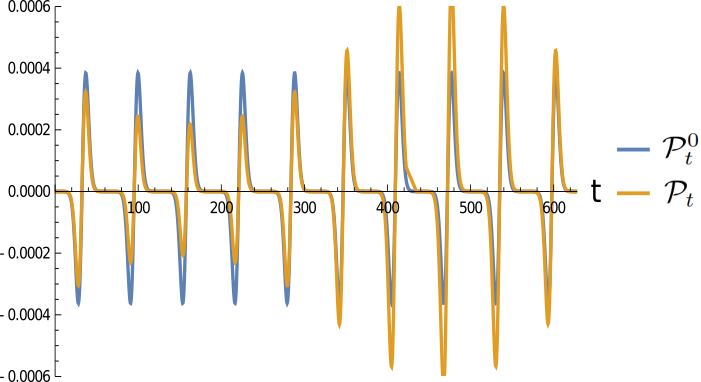}
    \caption{${\cal P}_t^0$ and ${\cal P}_t$ for the 2LS defined in  \eqref{dt}--\eqref{eq_rates_har} with $A=2$, $\omega_0=0.1$, $\omega_\text{B}=0.01$, $\beta=6$, $\ve=0.1$ and $h=1$.}
\label{fig_harm_P0_P}
\end{figure}

Let the energy of the excited state in \eqref{rates} be given by
\begin{equation}\label{dt}
  D_t = A (1 + \frac{h}{2} \cos(\omega_0 t)),
\end{equation}
with $0\leq h< 2$, where $h=0$ is the equilibrium reference for energy difference $A$.  We choose transition rates
\begin{eqnarray}
\label{eq_rates_har}
  \alpha_t := k_t(g,e) &=&\frac{e^{-\beta_t D_t/2}}{2\cosh(\beta_t D_t/2)},  \nonumber \\ 
  \delta_t := k_t(e,g) &=&\frac{e^{\beta_t D_t/2}}{2\cosh(\beta_t D_t/2)}. 
\end{eqnarray}
Comparing with \eqref{rates}, we have taken reactivities
\[
\nu_t^0 = \left(2\cosh(\beta D_t/2)\right)^{-1}
\]
In that way, the transition rates remain bounded as $\beta\uparrow \infty$ which is physically reasonable.\\
The Master equation \eqref{me} simplifies to
\begin{equation}\label{men}
    \partial_t \rho_t(e) +\rho_t(e) =  \alpha_t
\end{equation}
We have calculated \eqref{power} and \eqref{power1}. The (constant temperature) dissipated power ${\cal P}_t^0$ and the (varying temperature) dissipated power ${\cal P}_t$ are shown in Fig.~\ref{fig_harm_P0_P}. While ${\cal P}_t^0$ is periodic with frequency $\omega_0$, ${\cal P}_t$ need not be periodic unless $\omega_0$ and $\omega_\text{B}$ are commensurable.

\begin{figure}[h]
    \centering
    \includegraphics[width=0.7\textwidth]{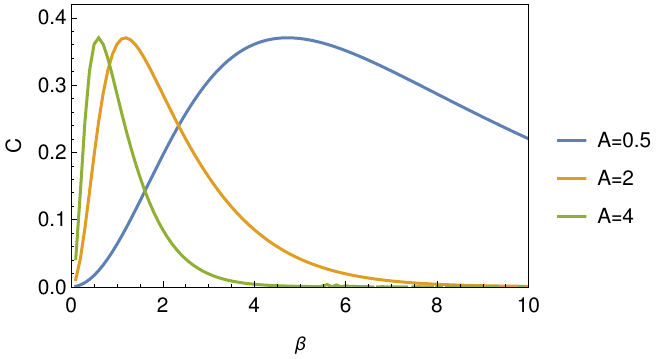}
    \caption{Nonequilibrium heat capacity $C$ {\it vs} $\beta$ for $h=1$, with $A=0.5,2,4$, and with $\omega_0=0.2$, $\omega_\text{B}=0.01$.}
\label{fig_harm_C_Eo}
\end{figure}

We next apply the procedure of \eqref{eq_C} for ${\cal P}_t^1 = ({\cal P}_t - {\cal P}^0_t)/\varepsilon$.  We have checked that it suffices to take the temperature modulation at a small frequency $\omega_\text{B}$ which is at least about 10 times smaller than $\omega_0$.  In  
Fig.~\ref{fig_harm_C_Eo} we see the heat capacity (always with $k_B=1$) as a function of inverse temperature $\beta$ for different values of the reference amplitude $A$.

\begin{figure}[H]
     \centering
     \begin{subfigure}{0.49\textwidth}
         \centering
         \def\svgwidth{0.8\linewidth}        
         \includegraphics[scale = 0.84]{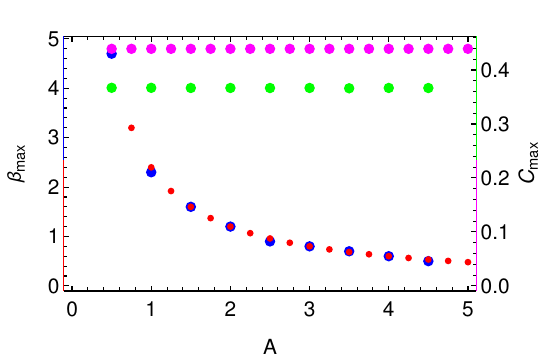}
         \caption{}
     \end{subfigure}
     \hfill
     \begin{subfigure}{0.49\textwidth}
         \centering
         \def\svgwidth{0.8\linewidth}        
         \includegraphics[scale = 0.75]{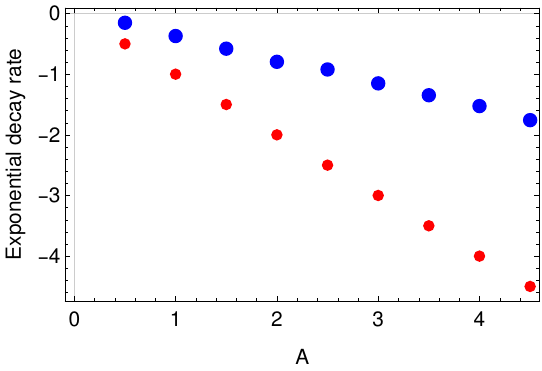}
         \caption{}
     \end{subfigure}
        \caption{(a) Position $\beta_\text{max}$ of the maximum of the heat capacity (in blue for left vertical axis), and the peak height $C_\text{max}$ (in green for right vertical axis) {\it vs} $A$ for $\omega_0=0.1$, $\omega_\text{B}=0.01$ and $h=1$. The equilibrium reference is indicated in red (for  $\beta_\text{max}$) and in magenta (for  $C_\text{max}$). (b) The slope of the exponential decay (for large $\beta$) versus $A$ in blue for the system with $\omega_0=0.1$, $\omega_\text{B}=0.01$ and $h=1$, and in red for an equilibrium system. The decay rate is $2/5$, and is the same for other frequencies $\omega_0=0.2$ and $\omega_0=0.05$.}
\label{fig_harm_C_max_y}
\end{figure}

Fig.~\ref{fig_harm_C_max_y}(a) shows the position $\beta_\text{max}$ of the maximum of the heat capacity.  It behaves as $\beta_\text{max} \approx 2.4/A$ as is the case for the Schottky anomaly for equilibrium 2LS as well. The Schottky peak (height $C_\text{max}$) is about constant for different values of $A$, and slightly lower than the equilibrium reference. 
That relates to the decay--rate of the heat capacity at low temperatures. In Fig.~\ref{fig_harm_C_max_y}(b) we see that the rate of the exponential decay for large $\beta$ (small temperature) is about $2/5$  of the decay rate in equilibrium.

\begin{figure}[H]
     \centering
     \begin{subfigure}{0.49\textwidth}
         \centering
         \def\svgwidth{0.8\linewidth}        
         \includegraphics[scale = 0.84]{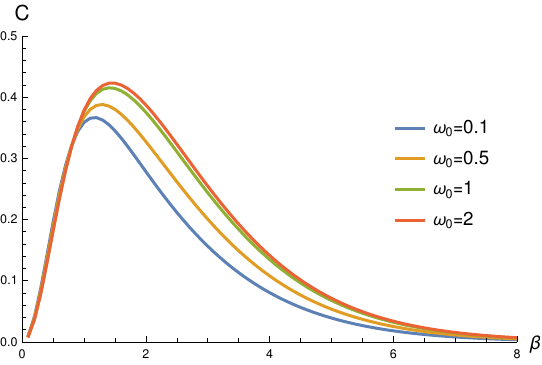}
         \caption{}
     \end{subfigure}
     \hfill
     \begin{subfigure}{0.49\textwidth}
         \centering
         \def\svgwidth{0.8\linewidth}        
         \includegraphics[scale = 0.84]{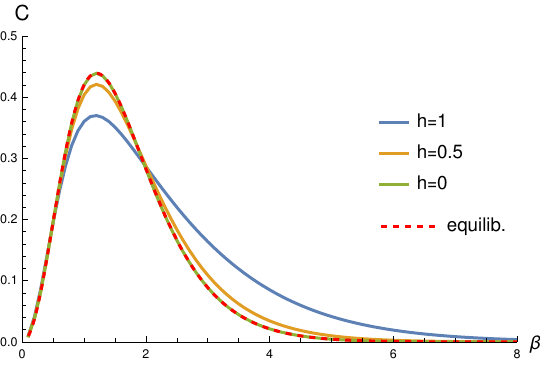}
         \caption{}
     \end{subfigure}
        \caption{(a) $C$ {\it vs} $\beta$ for different values of $\omega_0$, with $A=2$, $h=1$ and $\omega_\text{B}=0.01$. (b) $C$ {\it vs} $\beta$ for different values of $h$, with $A=2$, $\omega_0=0.2$ and $\omega_\text{B}=0.01$.  The case $h=0$ coincides with the equilibrium value \eqref{eq_C_eq}.}
\label{fig_harm_C_h_2}
\end{figure}

In Fig.~\ref{fig_harm_C_h_2}(a) we see how the heat capacity changes with different values of the frequency $\omega_0$. The differences are most outspoken at the low--temperature side of the Schottky peak and saturate for large driving frequencies. 
When the parameter $h \to 0$ the system goes to equilibrium, with heat capacity
\begin{equation}
\label{eq_C_eq}
    C_{eq} = \frac{(A \beta)^2}{4 \cosh^2(A \beta /2)},
\end{equation}
as is plotted in Fig.~\ref{fig_harm_C_h_2}(b) as well.

\subsection{Stroboscopic variation}
\label{strob}

\begin{figure}[h]
    \centering
    \includegraphics[width=0.35\textwidth]{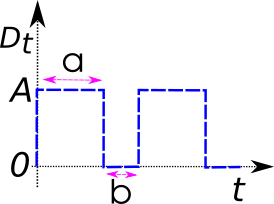}
    \caption{Stroboscopic energy variation in a 2LS.}
\label{fig_strob}
\end{figure}

Next we consider a 2LS with energies $E(g)=0$ for the ground state and $E(e)=D_t$ for the excited state, where
\begin{equation}\label{str}
D_t=
    \begin{cases}
        A & \text{if } \text{mod}(t,a+b) < a,\\
        0 & \text{otherwise}, 
    \end{cases}
\end{equation}
Here, mod$(t,\cal{T})$ gives the remainder of $t$ divided by $\cal{T}$, and $\cal{T}$ $=a+b$ is the period of the stroboscopic variation shown in Fig~\ref{fig_strob}.\\
The transition rates are given by
\begin{eqnarray}
  \alpha_t:= k_t(g,e) &=&\frac{e^{-\beta_t D_t/2}}{2\cosh(\beta_t D_t/2)},\nonumber  \\
  \delta_t:= k_t(e,g) &=&\frac{e^{\beta_t D_t/2}}{2\cosh(\beta_t D_t/2)}. 
\end{eqnarray}
The steady distribution \eqref{rh} can be obtained explicitly because of the simplicity of \eqref{str}.  It leads to the dissipated power(s) (from \eqref{power}--\eqref{power1}) and the heat capacity  (from Eq.~(\ref{eq_C}).\\

\begin{figure}[H]
     \centering
     \begin{subfigure}{0.49\textwidth}
         \centering
         \def\svgwidth{0.8\linewidth}        
         \includegraphics[scale = 0.84]{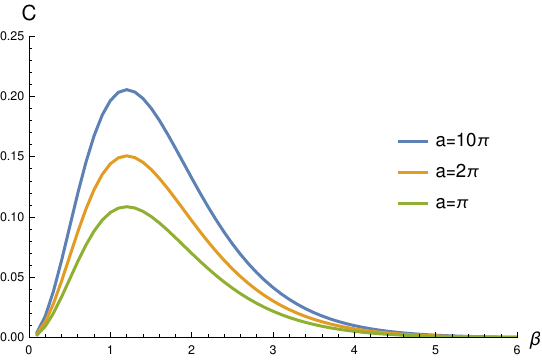}
         \caption{}
     \end{subfigure}
     \hfill
     \begin{subfigure}{0.49\textwidth}
         \centering
         \def\svgwidth{0.8\linewidth}        
         \includegraphics[scale = 0.84]{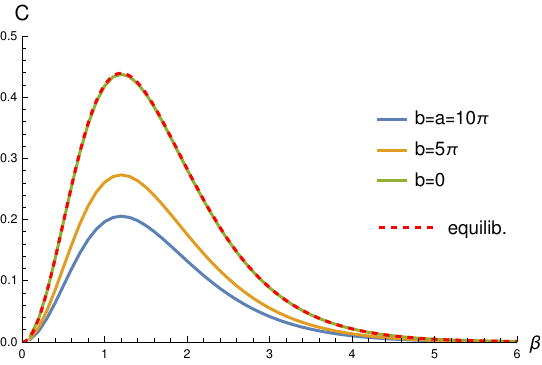}
         \caption{}
     \end{subfigure}
        \caption{(a) $C$ {\it vs} $\beta$ for  $a=b$ with different values of $a$, and with $A=2$ and $\omega_\text{B}=0.01$. With $\omega_0=2 \pi /\cal{T}$, the values $a=10 \pi, 2 \pi, \pi$ correspond to $\omega_0=0.1, 0.5, 1$, respectively. 
        (b) $C$ {\it vs} $\beta$ for the stroboscopic 2LS with fixed $a=10\pi$ and for different values of $b$. The values $b=10\pi,5\pi,0$ correspond to $\omega_0=0.1, 0.1\overline{3},0.2$, respectively. The equilibrium heat capacity \eqref{eq_C_eq} is plotted in dashed line. Parameters used: $A=2$ and $\omega_\text{B}=0.01$.}
\label{fig_strob_a}
\end{figure}

In Fig.~\ref{fig_strob_a}(a) is shown how the heat capacity changes with different values of the period $\cal{T}$ for the symmetric case $a=b$.  Fig.~\ref{fig_strob_a}(b) gives asymmetric cases.
On the other hand, for $b=0$, {\it i.e}, $D_t=A$ for all $t$,
the system is in equilibrium, as verified in Fig.~\ref{fig_strob_a}(b).

We note that the heat capacities for stroboscopic variation of the energy gap are different from the case of harmonic variation at the same frequency and amplitude.

\section{2LS with a kinetic barrier}
\label{barrier}
\begin{figure}[h]
    \centering
    \includegraphics[width=0.5\textwidth]{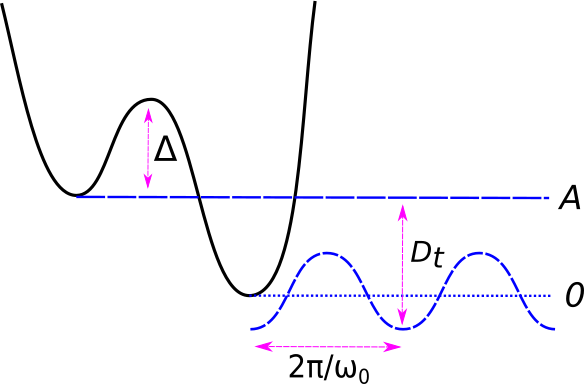}
    \caption{2LS with a kinetic barrier $\Delta$. The energy difference $D_t$ changes harmonically in time.}
\label{fig_barrier_scheme}
\end{figure}

We consider here a 2LS with energies $E(g)=A \, h \cos(\omega_0 t)/2$ for the ground state and $E(e)=A$ for the excited state, and with a barrier $\Delta$ between the two states; see Fig.~\ref{fig_barrier_scheme}. The energy difference is  $D_t=A (1- h \cos(\omega_0 t)/2)$. We have computed the heat capacity $C$ in the same way as above, by quasistatic variation of the temperature.  More specifically, we choose transition rates
\begin{eqnarray}
  \alpha_t := k_t(g,e) &=& e^{-\beta_t (D_t+\Delta)},\,  \nonumber \\
  \delta_t := k_t(e,g) &=&e^{-\beta_t \Delta}.
\end{eqnarray}
where the inverse temperature follows \eqref{timebeta}.

\begin{figure}[H]
     \centering
     \begin{subfigure}{0.49\textwidth}
         \centering
         \def\svgwidth{0.8\linewidth}        
         \includegraphics[scale = 0.84]{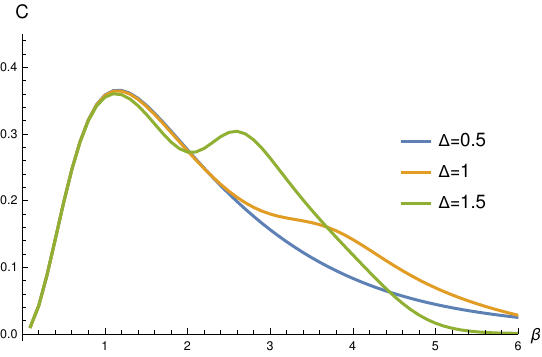}
         \caption{}
     \end{subfigure}
     \hfill
     \begin{subfigure}{0.49\textwidth}
         \centering
         \def\svgwidth{0.8\linewidth}        
         \includegraphics[scale = 0.84]{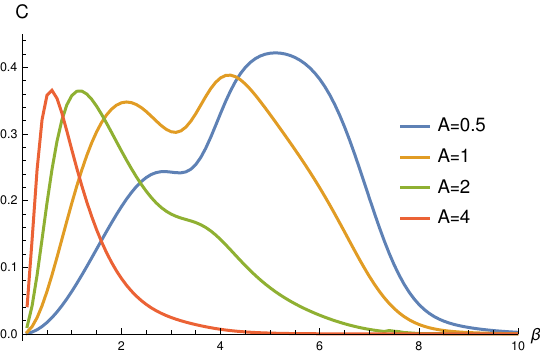}
         \caption{}
     \end{subfigure}
        \caption{$C$ {\it vs}. $\beta$ for the 2LS in Fig.~\ref{fig_barrier_scheme} with $h=1$, $\omega_0=0.03$ and $\omega_\text{B}=0.001$; (a) for different values of $\Delta$,  with $A=2$, (b) for different values of $A$ with $\Delta=1$. }
\label{fig_barrier_delta}
\end{figure}

Figs.~\ref{fig_barrier_delta} show the heat capacity for the system displayed in Fig.~\ref{fig_barrier_scheme} as a function of $\beta$ for different values of the barrier $\Delta$ and different amplitudes $A$.  We see an extra shoulder indicating the presence of  $\Delta$. From fitting, we observe that the position of that second peak in the heat capacity (in Fig.~\ref{fig_barrier_delta}(a)), as a function of $\beta$,  moves as $3.7/\Delta$.   It means that heat capacity measurements are now able to locate the presence of a kinetic barrier; see \cite{activePritha, jchemphys} for examples in the stationary case.  Its position and height also change with $A$; see  Fig.~\ref{fig_barrier_delta}(b).\\

\begin{figure}[H]
     \centering
     \begin{subfigure}{0.49\textwidth}
         \centering
         \def\svgwidth{0.8\linewidth}        
         \includegraphics[scale = 0.84]{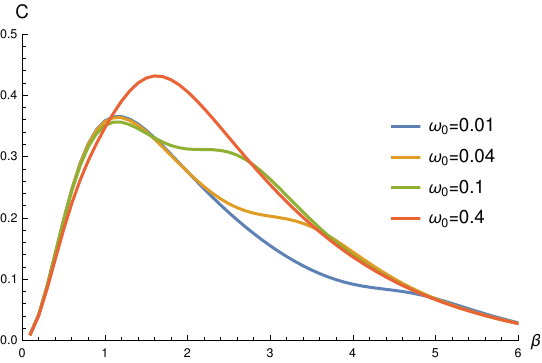}
         \caption{}
     \end{subfigure}
     \hfill
     \begin{subfigure}{0.49\textwidth}
         \centering
         \def\svgwidth{0.8\linewidth}        
         \includegraphics[scale = 0.84]{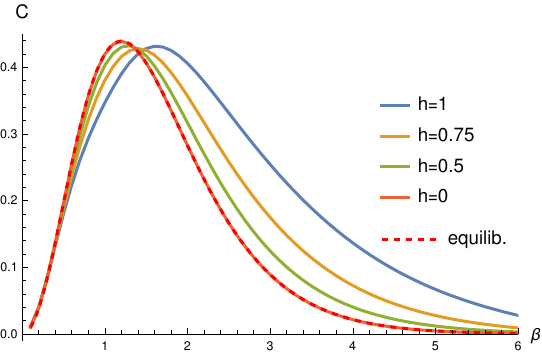}
         \caption{}
     \end{subfigure}
        \caption{$C$ {\it vs} $\beta$ , for the 2LS in Fig.~\ref{fig_barrier_scheme} with  $\Delta=1$, $A=2$  and $\omega_\text{B}=0.001$; (a)  for different values of $\omega_0$, and at $h=1$; (b) for different values of $h$ at $\omega_0=0.4$. Also, the equilibrium heat capacity (coinciding with the case where $h=0$) is shown for comparison. }
\label{fig_barrier_w0}
\end{figure}

Figs.~\ref{fig_barrier_w0} give $C$ as a function of $\beta$ for different values of $\omega_0$, and of the nonequilibrium parameter $h$.  Again, for larger frequencies and for smaller $h$, the equilibrium curve is approximated.

\section{Conclusions and outlook}

We have reported a theoretical study of the quasistatic thermal response of systems subject to time--dependent forces. The 2LS, {\it e.g.}, in the form of a time--dependent double well, provides a proof--of--principle for steady state calorimetry by associating a heat capacity to an open periodically--driven system in contact with a heat bath, even at room temperature. We have given its dependence on driving frequency, driving amplitude, and bath temperature for various dynamics of a two--level system (in the incoherent approximation). Interesting features include the dependence of that heat capacity on kinetic features (such as a barrier and choice of transition rates) and the slower decay to zero at vanishing temperatures.

Limitations to the given approach are mostly the computational complexity and, experimentally, the care that must be given in distinguishing the steady dissipated power ${\cal P}^0$ from the temperature--modulated $\cal P$. One other difficulty is to deal with a necessarily open system on which work is being performed and to screen it at the same time from other influences. Experimental elaboration is in progress and will face the natural problems of pioneering work.

The analysis is a stepping stone for similar studies of calorimetry in periodically driven systems. As one specific example, the theoretical derivation or understanding of the heat capacity curves for ferroelectric materials undergoing a hysteresis loop, \cite{cera}, is still lacking largely. We also hope that this work may stimulate experimental work on nonequilibrium calorimetry as it promises to open new windows on the functioning of driven materials.


\bibliographystyle{unsrt}  
\bibliography{chr}
\onecolumngrid

\end{document}